\documentclass[aps,prl,twocolumn,superscriptaddress,nofootinbib,longbibliography]{revtex4-2}
\pdfoutput=1

\usepackage{graphicx}
\usepackage{amsmath,amssymb,amsfonts}
\usepackage[colorlinks,citecolor=blue,linkcolor=blue,urlcolor=blue, breaklinks=true]{hyperref}
\usepackage{color}
\usepackage[normalem]{ulem}

\usepackage[T1]{fontenc}

\usepackage{physics} 
\usepackage[percent]{overpic} 
\usepackage{bm}

\usepackage[english]{babel}

\usepackage{graphicx}

\usepackage{xspace}

\newcommand{\munu}{{\mu\nu}\xspace}

\begin{document}

\title{Locally mediated entanglement in linearised quantum gravity}

\author{Marios Christodoulou}
\affiliation{Institute for Quantum Optics and Quantum Information (IQOQI) Vienna, Austrian Academy of Sciences, Boltzmanngasse 3, A-1090 Vienna, Austria}
\affiliation{Vienna Center for Quantum Science and Technology (VCQ), Faculty of Physics, University of Vienna, Boltzmanngasse 5, A-1090 Vienna, Austria}

\author{Andrea {Di Biagio}}
\affiliation{Institute for Quantum Optics and Quantum Information (IQOQI) Vienna, Austrian Academy of Sciences, Boltzmanngasse 3, A-1090 Vienna, Austria}
\affiliation{Dipartimento di Fisica, Sapienza Universit\`a di Roma, Piazzale Aldo Moro 5, Roma, Italy}

\author{Markus Aspelmeyer}
\affiliation{Institute for Quantum Optics and Quantum Information (IQOQI) Vienna, Austrian Academy of Sciences, Boltzmanngasse 3, A-1090 Vienna, Austria}
\affiliation{Vienna Center for Quantum Science and Technology (VCQ), Faculty of Physics, University of Vienna, Boltzmanngasse 5, A-1090 Vienna, Austria}
\affiliation{Research Platform TURIS, University of Vienna, Vienna, Austria.}

\author{\v Caslav Brukner}
\affiliation{Institute for Quantum Optics and Quantum Information (IQOQI) Vienna, Austrian Academy of Sciences, Boltzmanngasse 3, A-1090 Vienna, Austria}
\affiliation{Vienna Center for Quantum Science and Technology (VCQ), Faculty of Physics, University of Vienna, Boltzmanngasse 5, A-1090 Vienna, Austria}
\affiliation{Research Platform TURIS, University of Vienna, Vienna, Austria.}

\author{Carlo Rovelli}
\affiliation{Aix-Marseille University, Universit\'e de Toulon, CPT-CNRS, Marseille, France,}
\affiliation{Department of Philosophy and the Rotman Institute of Philosophy, Western University, London ON, Canada,}
\affiliation{ Perimeter Institute, 31 Caroline Street N, Waterloo ON, Canada} 

\author{Richard Howl}
\affiliation{Quantum Group, Department of Computer Science, University of Oxford, Wolfson Building, Parks Road, Oxford, OX1 3QD, United Kingdom}
\affiliation{QICI Quantum Information and Computation Initiative, Department of Computer Science, The University of Hong Kong, Pokfulam Road, Hong Kong}

\date{\small\today}

\begin{abstract} \noindent 
The current interest in laboratory detection of entanglement mediated by gravity was sparked by an information--theoretic argument: entanglement  mediated by a local field certifies that the field is not  classical.  Previous derivations of the effect modelled gravity as instantaneous; here we derive it from linearised quantum general relativity while keeping Lorentz invariance explicit, using the path integral formalism. In this framework,  entanglement is clearly mediated by a quantum feature of the   field.
We also point out the possibility of observing \emph{retarded}  entanglement, which cannot be explained by an instantaneous interaction. This is a difficult experiment for gravity, but is plausible for the analogous electromagnetic case.
\end{abstract}

\maketitle 

It is often assumed that quantum gravitational effects only show up at  high--energy or short length scale regimes, out of reach of current technology.  Recent proposals for low--energy table--top experiments  could be game changers  \cite{pikovski2012probing,bose2017spin,marletto2017gravitationallyinduced,krisnanda2017revealing,krisnanda2020observable,*albalushi2018optomechanical,*weiss2021large,howl2021nongaussianity, christodoulou2020experiment,*christodoulou2020possibility}. Rapid technological progress in quantum manipulation of solid-state matter at larger microscopic mass scales \cite{magrini2021realtime,*tebbenjohanns2021quantum,delic2020cooling}, and in gravitational measurements at smaller mesoscopic mass scales \cite{westphal2021measurement}, have raised expectations that probing gravitational phenomena of quantum source masses may be within reach \cite{rovelli2021considerations}. In particular, it might be possible to detect entanglement between two masses generated by their gravitational interaction, or GIE  (Gravity Induced  Entanglement)\cite{bose2017spin,marletto2017gravitationallyinduced}.  

Verifying GIE would spectacularly support what is expected from most  tentative quantum gravity theories: spacetime has quantum properties. It would also falsify---or put limits on---the alternatives that have been considered in the absence of empirical evidence for quantum gravity: for example, that gravity is a classical field obeying semiclassical Einstein equations \cite{carlip2008quantum,oppenheim2018postquantum,kafri2014classical}, or  that quantum mechanics breaks down at a scale before measurable quantum gravity effects appear \cite{penrose1996gravity,diosi1989models, bassi2013models}. Specifically, a general quantum information argument has been invoked to argue that GIE would rule out the possibility that the gravitational field is a local, classical field    \cite{bose2017spin,marletto2017gravitationallyinduced,marletto2018when,galley2021nogoa,pal2021experimental,marletto2020witnessing,galley2021nogoa}. The argument is based on the fact that local operations and classical communication (LOCC) cannot produce  entanglement according to quantum theory \cite{horodecki2009quantum}, as well as to more general approaches \cite{galley2021nogoa,marletto2020witnessing}. Then, the argument goes, observing GIE  certifies that gravity cannot be described by classical physics: either the interaction is nonlocal, or it is non-classical.

However, the implications of GIE detection are being debated even when assuming linearised quantum gravity. In this context, some claims have been made that the experiment does not detect a quantum property of the gravitational field \cite{anastopoulos2018comment,anastopoulos2021gravitational}. The disagreement partially stems from the fact that the effect has generally been computed within the approximation of an instantaneous interaction. Indeed, since the imagined experiment involves masses with non--relativistic motion placed close to each other, in this regime gravity can effectively be described without the need of a dynamical field. But this approach hides a core ingredient of the theory: relativistic locality.  There is strong independent experimental evidence that the gravitational interaction is not instantaneous.

We provide a derivation of the effect within linearised quantum gravity, using the path--integral formalism, which keeps the symmetries explicit. In particular, spacetime locality is kept manifest. Starting from two established paradigms of physics\footnote{
 Some investigations of the black hole information paradox hypothesize that the ultraviolet sector of quantum gravity may have important low-energy implications, including  generating long-distance entanglement - see e.g. \cite{berglund2022infrared} and references therein.  We doubt this possibility is relevant for the phenomenon at hand, but we do not address this question further. 
}---general relativity and quantum field theory---we show here that the quantum phases responsible for gravity mediated entanglement production are on--shell actions (cf.~Eq.~\eqref{eq:phases}), which we compute below  (cf.~Eq.~\eqref{eq:SosGravity}). This provides an explicitly Lorentz invariant, hence spacetime local,  and gauge invariant description of GIE. This is our main result.

In our analysis, GIE turns out to be due to the fact that the overall path integral reduces to a finite sum, in each term of which the functional integral can be estimated by a `semiclassical' saddle point approximation. In other words, in this formalism the effect is due to a genuinely quantum feature of the gravitational field: the possibility to be  in a quantum superposition of distinct semiclassical configurations.

This implies that, in the context of linearised quantum gravity, GIE arises due to a quantum superposition of spacetimes  \cite{christodoulou2019possibility}, each propagating information causally. Thus, information travels in a quantum superposition of wavefronts in the field and entanglement starts being generated only \emph{after a light crossing time has elapsed}.

We consider a consequence of this local propagation in linearised quantum gravity: the existence of an experiment where both entanglement and relativistic locality can be observed, thus, incompatible with an instantaneous interaction description.
For gravity, this is currently out of reach but we find
the analogous experiment in electromagnetism to be feasible. Since our analysis proceeds completely analogously for the electromagnetic case, this would inform the outcome of the gravitational experiment.

\subsection*{Locally Mediated Entanglement from the Path Integral of the Quantum Field}

Consider the experimental setup in \cite{bose2017spin} that comprises two\footnote{The formulas are the same for an arbitrary number of particles.} masses $m_a$ ($a=1,2$), each with an embedded spin--1/2 degree of freedom.  At time $t^i$, the particles are at initial positions $x_a^i$ and are then put in a spin--dependent planar motion $x^{s_a}_a(t)$, by being passed through inhomogeneous and possibly time varying magnetic fields $B_z$ oriented along the axis $z$, perpendicular to the plane of motion. We denote $\ket{\sigma}=\otimes_a \ket{s_a}$ the spin configurations, where $s_a \in \{\uparrow, \downarrow\}$.
 
The spacetime curvature is assumed to be small, and so  the linear approximation of general relativity holds. We denote the gravitational perturbation sourced by the particles as $\mathcal{F}$.\footnote{For electromagnetism, $\mathcal{F}$ is the four--potential and for gravity is the perturbation of the metric.} Preparing each of the particles in a spin superposition state, the magnetic field $B_z$ drives the particles into a path--superposition by coupling to the spins $s_a$. The field $\mathcal{F}$ couples to the masses $m_a$ of the moving particles. After recombining the interferometer paths at time $t_2$, see Fig.\! \ref{fig:ST},  a spin measurement is performed on each particle at time $t^f$. The spins can become entangled due to the gravitational interaction between the masses $m_a$. The coupling of $B_z$ with  $\mathcal{F}$, the backreaction of $s_a$ on $B_z$, and the backreaction of $\mathcal{F}$ on the particle trajectories $x^{s_a}_a(t)$, are taken to be negligible. 
 
 

The transition amplitudes are computed using the path integral 
\begin{equation}
\label{eq:partFun}
\int \! \mathcal{D} \mathcal{F'} \mathcal{D} x' \, \exp\left(\frac{iS}{\hbar}\right)
\end{equation}
where $S=\left[ x'_a(t),\mathcal{F'}(x,t);m_a,B_z, \sigma \right]$;  $\mathcal{D} x'=\prod_a \mathcal{D} x'_a $, and the integration is over field configurations $\mathcal{F'}(x,t)$ and paths of the particles $x'_a(t)$.\footnote{We assume the apparatus very massive and its gravitational field approximately homogeneous in the relevant region. See the Supplementary Material.} The quantities $m_a$, $B_z$ and $\sigma$ are not affected by the dynamics in  \eqref{eq:partFun} .


The path that each particle takes is determined by the spin, which does not change along the path. Thus, the joint evolution is of the form
\begin{equation}
\label{eq:evolutionOp}
U_{i\rightarrow f}  = \sum_{\sigma}\ketbra{\sigma}\otimes U^{\sigma}_{i\rightarrow f}, 
\end{equation}
with $U^\sigma_{i\rightarrow f} $ defined by folding \eqref{eq:partFun} with initial and final states
$\ket{\psi^{i,f}}= \ket{{\mathcal F}^{i,f}[x_a^{i,f}]} \otimes \ket{x_a^{i,f}}$, where the paths and field states are assumed pure and separable at $t^{i,f}$. The boundary conditions are taken the same for all spin configurations $\sigma$. The time $t^f$ is far enough in the future for the field to have time to relax in the vicinity of the spin measurement. It is sufficient to take the boundary conditions as given by the static Newtonian field ${\mathcal F}^{i,f}[x_a^{i,f}]$ of masses sitting at the initial and final particle positions $x_a^{i,f}$.

The task  is to calculate $U^{\sigma}_{i\rightarrow f}$ up to normalisation. The field integration can be heuristically performed by  a stationary phase approximation, keeping the contribution of the field configurations $\mathcal{F}[x_a(t)]$ that solve the classical field equations sourced by particles of mass $m_a$ with classical paths $x_a(t)$ and boundary conditions $\ket{\psi^{i,f}}$. Then,
\begin{equation}
\label{eq:QMpathintegral}
U^\sigma_{i \rightarrow f} \propto \int_i^f \! \mathcal{D} x' \, \exp\left(\frac{iS\big[x'_a,\mathcal{F}[x'_a] \big]}{\hbar}\right) \ketbra{\psi^f}{\psi^i} .
\end{equation}
This approximation allows us to sidestep the rigorous definition of the path--integral 
\cite{burgess2004quantum,wallace2021quantum} and neglects quantum fluctuations.

Between times $t^i$ and $t^f$, for each spin configuration $\sigma$ there is a classical path $x_a^{s_a}$ determined by the magnetic field $B_z$ coupled to the spin $s_a$ of each particle. These paths can be taken as orthogonal states, and the remaining integral approximated by a second stationary phase approximation, keeping only the contribution on these paths
\begin{equation}\label{eq:QMpathintegral2}
 U^{\sigma}_{i\rightarrow f} \propto \exp\left(\frac{iS^\sigma\big[x_a^{s_a},\mathcal{F}[x_a^{s_a} ]\big]}{\hbar}\right) \ketbra{\psi^f}{\psi^i} ,
\end{equation}
 Here, for a given spin configuration $\sigma$, $S^\sigma$ is the \emph{on--shell action for the joint system of spins, paths and field}.

The action $S$ splits as $S= S_0+ S_\mathcal{F}$. $S_0$ does not depend on $\mathcal{F}$, it contains the matter kinetic terms and the coupling of $B_z$ with the spins $s_a$. $S_0$ can be calculated, or measured, separately. For simplicity, we assume the setup to be chosen so that $S_0$ is the same for all $\sigma$ and becomes a global phase. $S_\mathcal{F}$ contains the on--shell contributions of the kinetic terms for the field $\mathcal{F}$ and of the coupling of $\mathcal{F}$ with the masses $m_a$ along their motion $x_a$: $S_\mathcal{F}$ contains the \emph{field mediation}. We define 
\begin{equation}
\label{eq:phases}
\phi_\sigma =\frac{S^\sigma_{\mathcal{F}}\big[x_a^{s_a},\mathcal{F}[x_a^{s_a}] \big]}{\hbar}.
\end{equation}

Given an initially separable state
$\ket{\Psi^i} \propto \ket{\psi^{i}} \otimes  \sum_{\sigma} A_\sigma \ket{\sigma}$  of field, paths and spins, with $A_\sigma$ complex amplitudes, the final state is given by 
\begin{equation}
\label{eq:fState}
  \ket{\Psi^f} =  U_{i\rightarrow f}  \ket{\Psi^i} \propto  \ket{\psi^f} \otimes \sum_{\sigma}  A_\sigma \, e^{i \phi_\sigma} \ket{\sigma}.  
\end{equation}
Note that boundary states are not entangled with the spin configurations at initial and final times. However, depending on the values of $S^\sigma_\mathcal{F}$, entanglement can be produced among the spin degrees of freedom. The phases $\phi_\sigma $ are the result of the entanglement production \emph{mediated through $\mathcal{F}$} \cite{bose2017spin,marletto2017gravitationallyinduced,christodoulou2019possibility}. We have shown that the phases $\phi_\sigma $ are on--shell actions, therefore they are \emph{manifestly local and gauge invariant}. Differences of $\phi_\sigma $ for different $\sigma$, the relative phases among branches, are the observables measured by the experiment. We now compute $\phi_\sigma $. 


\subsection*{Covariant phases for the gravitational field of moving particles}\label{sec:on-shell-action}

The action of linearized gravity coupled to matter is gauge invariant.  On--shell, it reads
\begin{align} \label{SLorentzG}
	S_\mathcal{F} = \frac{1}{4} \int \dd^4 x \; h_{\mu \nu} T^{\mu \nu}
\end{align}
where $h_{\mu \nu}$ is the metric perturbation and $T_{\mu\nu}$ the energy--momentum tensor. Modelling the masses as point particles with arbitrary timelike trajectories, their gravitational field is the gravitational analogue of the Li\'{e}nard–Wiechert potentials of electromagnetism \cite{kopeikin1999lorentz,jackson1999classical,Supplementary}. The on--shell action  \eqref{SLorentzG} is then given by
\begin{align}\label{eq:SosGravity}
	S_{\mathcal{F}} & = \frac{G}{c^4 } \sum_{a,b}^{a \neq b}  \int \dd t   \frac{m_a m_b \bar{V}^{\mu \nu}_a(t_{ab}) V_{b \mu \nu} (t)}{d_{ab}(t) - \bm{d}_{ab}(t) \cdot \bm{v}_a(t_{ab})/c},
\end{align}	
where
$V^{\mu \nu}_a = \gamma_a v_{a}^{\mu}v_{a}^{ \nu}$,
$v^{\mu}_a = (c, \bm{v}_a )$,
$\bm{v}_a $ the three velocity,
and $\gamma_a$ the Lorentz factor. 

The analogous formula in electromagnetism is obtained by replacing $\bar{V}^{a\mu \nu} V_{b\mu \nu} \rightarrow v^{ \mu}_a v_{b \mu}$, $ m \rightarrow q$ and $G/c^4 \rightarrow \kappa_e/2c^2$ where $q$ is the charge and $\kappa_e$ Coulomb's constant. For the notation and a detailed derivation of \eqref{eq:SosGravity}  see the Appendix and the Supplementary Material. The crucial point is that the distance $d_{ab}$ and the time $t_{ab}$ are \textit{retarded} quantities.

The action \eqref{eq:SosGravity} is a sum of two terms per pair of particles. Each term is the contribution from one particle at coordinate time $t$ interacting with the other causally, that is, with retardation. The causal interaction between matter and the gravitational field is thus entirely encoded in $S_\mathcal{F}$. This manifestly Lorentz and gauge invariant quantity gives the observables measured in the experiment.

\subsection*{Slow--motion  approximation versus Newtonian limit}

When the source is `slow moving', meaning moving at non--relativistic speeds $\vert \bm{v}_a \vert \ll c$, we have 
$\bar{V}^{\mu \nu}_a(t_{ab}) V_{b \mu \nu}(t)  = c^4 + \mathcal{O}\left(c^3\vert \bm{v}_a \vert \right)$ and \eqref{eq:SosGravity} approximates
\begin{align} \label{Sint00}
	S^\sigma_{\mathcal{F}} \approx \frac{1}{2} G \sum_{a,b}^{a \neq b}  \int \! \dd t \, \frac{m_a m_b}{d^\sigma_{ab}(t)}.
\end{align}
In this  regime \emph{the interaction is still local}. The distance $d^\sigma_{ab}(t) = \vert \bm{x}^{s_b}_b(t)-\bm{x}^{s_a}_{a} (t^\sigma_{ab}) \vert$ depends on the \emph{retarded} time function $t^\sigma_{ab}(t)$.
While the speed of light $c$ has cancelled out in the prefactor of \eqref{Sint00}, it is still present implicitly in the definition of $t^\sigma_{ab}(t)$. Equation \eqref{Sint00} can be regarded as the causal version of Newton's law for gravitation. 

A different approximation for \eqref{eq:SosGravity} can be taken when the source's characteristic scale of time variation divided by $c$ is much larger than the distance of the source. Then, retardation in the field can be neglected in the vicinity of the source. This is a near--field approximation, it amounts to replacing the retarded time functions $t_{ab}$ in $\eqref{eq:SosGravity}$ with the coordinate time $t$, hence, modelling the gravitational interaction as an instantaneous interaction.
The slow--moving and near--field approximations do not imply each other, there are physical regimes when one is applicable and the other is not, and vice versa. When both approximations are applied, they yield the `Newtonian limit'. Taking in addition $d$ to be constant during a relevant time $T$, corresponding to considering a static approximation, we recover the formula used in the literature
\begin{align} \label{eq:Newtonian-action}
\phi_\sigma \approx \frac{ G m_1 m_2 T}{\hbar d}.
\end{align}
This expression for the phases $\phi_\sigma$ naively models an instantaneous interaction, but it is just an approximation to the manifestly local on--shell action \eqref{eq:SosGravity} of the joint system of paths, spins and field. 

\subsection*{Observable effect of retardation}

The effect of retardation can be quantified as the correction to the Newtonian limit \eqref{eq:Newtonian-action} by the slow--moving approximation \eqref{Sint00}. Importantly, a \textit{qualitatively} different behaviour can be observed when the spatial superposition of the particles happens entirely within spacelike separated regions.

\begin{figure}
    \centering
    \includegraphics[scale=0.18]{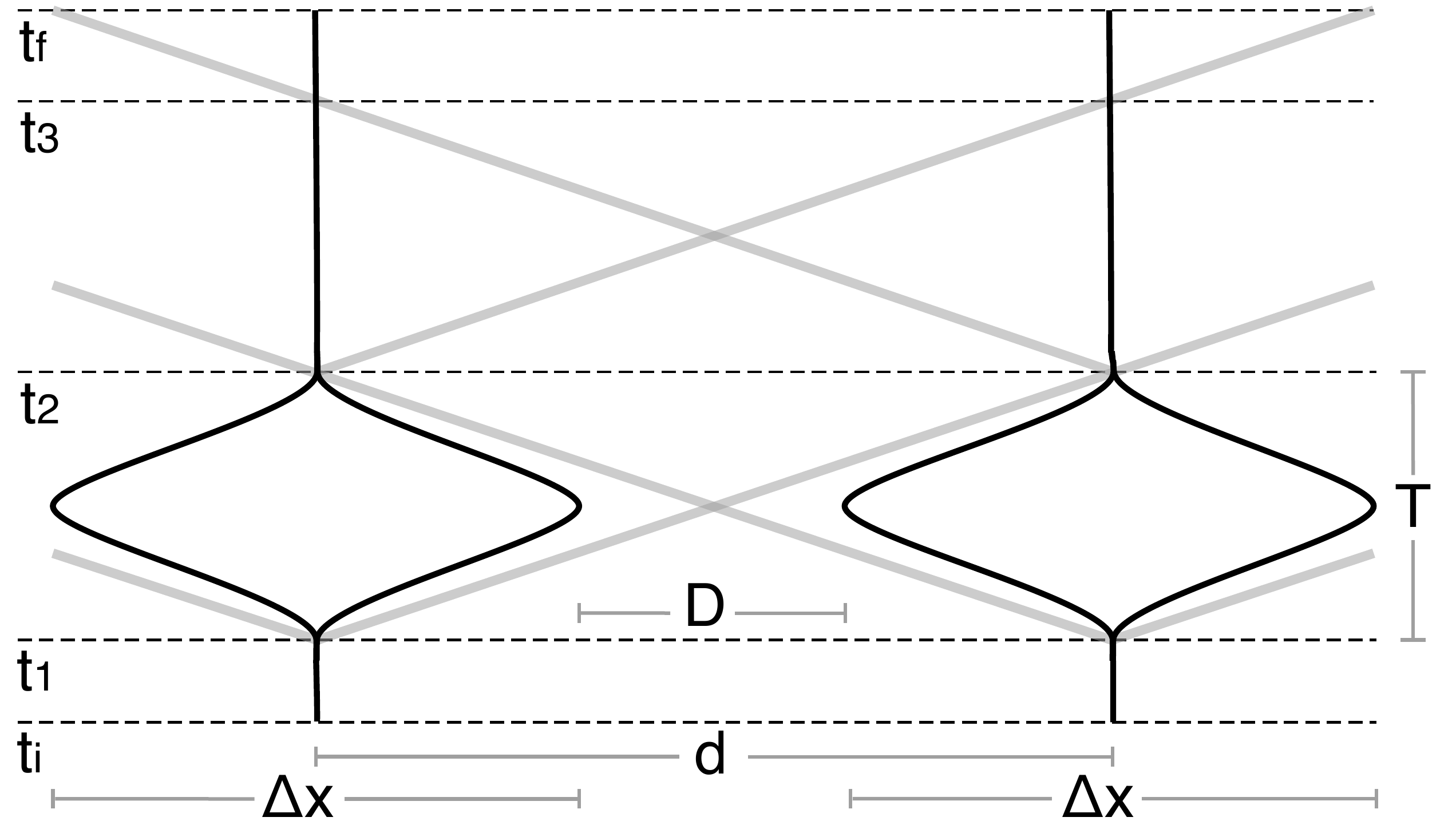}
    \caption{
    The lightcone structure forbids entanglement when the superposition happens at spacelike separation. This occurs when $d/T \geq c$.}
    \label{fig:ST}
\end{figure}

Take the particles at rest at a distance $d$ for all times $t<t_1$ and $t>t_2$. Between $t_1$ and $t_2$, the particles undergo a spin--dependent motion. The setup is such that
$c(t_2-t_1) < d$, so that the non--stationary parts of the worldlines are spacelike separated. From time
$t_3 = t_2+d/c$, the retarded position of each particle with respect to the other is again constant. See Figure \ref{fig:ST}. With this setup, no entanglement can be generated.

Let $x^{s_a}_a(t)$ be the displacement of particle $a$ from its initial position due to the coupling of the external magnetic field $B_z$ with its spin $s_a$ in the spin configuration $\sigma$. We remind that $\ket{\sigma}= \otimes_a \ket{s_a}$.
Using \eqref{Sint00}, $\phi_\sigma$ is a sum of integrals that can be done by splitting the domain of integration in four. Then,
\begin{equation}
\label{eq:split}
\begin{aligned}
 \int_{t^i}^{t^f}\frac{\dd t}{d_{21}^\sigma(t)} 
=& \int_{t^i}^{t_1}\frac{\dd t}{d} + \int_{t_1}^{t_2} \frac{\dd t}{d-x_1^{s_1}(t)} \\
&+  \int_{t_2}^{t_3}  \frac{\dd t}{d+x_2^{s_2}\big(t_{21}(t)\big)} + \int_{t_3}^{t^f}\frac{\dd t}{d}.
\end{aligned}
\end{equation}
Then, the phases are of the form
$\phi_\sigma = C + \phi_{s_a} + \phi_{s_b} + C'$ with terms that depend on at most one spin.%
\footnote{The same is true if we do not assume the slow-moving limit.} Thus, if the initial states of the spins is separable, so will be the final state. If, on the other hand, one calculates the phase in the Newtonian limit with instantaneous interaction, the spins result in an entangled state.

\subsection*{Experimental considerations}
\label{sec:gedanken}
 
The effect described above can in principle be observed experimentally, even though the parameters may be challenging. One possible way to achieve spacelike separation between the two interferometer loops of \cite{bose2017spin} is to increase the velocity $\bm{v}$ at which the particles traverse the apparatus. We denote $d$ the initial distance of the particles,  $\Delta x$ the maximum separation of the path superposition, and $D=d-\Delta x$ the minimum distance of the branches at closest approach, see Figure \ref{fig:ST}.
 
Using the Newtonian limit \eqref{eq:Newtonian-action}, and assuming $\Delta x\ll d$, the entanglement is maximal when \cite{bose2017spin}
$\Delta\phi \approx ({A}/{A_P})^2({\Delta x}/{d})^2({c T}/{D})=\pi$,
where $T=t_2-t_1$, $A$ is  the mass $m$ and $A_P$ is the Planck mass $m_P = \sqrt{\hbar c/G}$. For the Coulomb case, $A$ is the charge $q$ and $A_P$ is the Planck charge $q_P = \sqrt{4\pi\epsilon_0\hbar c}$.
 
As we showed above, when $d \geq c T $, no entangling interaction can take place between the particles. In other words: one can create a situation in which the Newtonian limit yields $\Delta\phi \sim 1$, while the actual value predicted by \eqref{eq:SosGravity} and \eqref{Sint00} is $\Delta\phi\sim 0$. Fixing the speed $v$ so that $d/T = c$ and assuming $\Delta x \ll d$ at these time-scales, achieving such maximum discrepancy would require $A\gg A_P$. For the gravity case this results in magnetic fields and coherence requirements that are not realistic for the foreseeable future (for comparison: current proposals operate in a regime $A \approx 10^{-10} A_P$ at much larger time scales, while the magnetic field requirements for coherent splitting scale with both mass and time).

Smaller effects of retardation are more easily measured. Let us assume, for the sake of the argument, that one can detect a one part in a thousand deviation from the Newtonian approximation.  One can estimate the retarded phases by replacing $T$ with $T-d/c$, which implies a correction $\delta(\Delta\phi)\approx({A}/{A_P})^2 ({\Delta x}/{d})^2$.
This will still require fairly large $A/A_P$, which is unlikely reachable for the gravity case. For the electromagnetic case however ion and electron interferometry offers a promising path \cite{hasselbach2009progress}.
For a single electron, $A/A_P\approx0.1$.
Assuming $d\approx1\,\mathrm{cm}$, $T\approx50\,\mathrm{ns}$ and that the superposition is produced by diffraction with a grating of periodicity $10\,\mathrm{nm}$, it is possible to produce $\Delta x/d \approx 0.3$ and thus the desired $\delta(\Delta\phi)\approx10^{-3}$. In an interferometer of length $10\,\mathrm{cm}$, this can be achieved with electron velocity of $v=10^{-2}c$, which is reachable in current electron microscopes. 

One possible way of testing for entanglement generation in such a scenario could be indirectly via controllable decoherence and recoherence of the single-electron interference signals \cite{kerker2020quantum}: if no entanglement is generated, both interferometers will show full (single-electron) coherence, while any generation of entanglement would decohere the single-electron interference signals. Both scenarios are accessible by changing the velocity of both beams. While this is not an easy experiment to perform, it is plausible for the near future. 

\subsection*{Discussion}

We considered experimental proposals aiming at observing the entanglement between two masses (or charges) due to the mediation of their gravitational (or electromagnetic) interaction. Entanglement happens because different quantum branches accumulate different phases. The phases were  previously computed using an instantaneous interaction, which  in part obscured the relevance of the experiment. 

We computed the phases from first principles and shown they are differences in on--shell actions. They are manifestly Lorentz invariant, hence causal, and gauge invariant. We considered the approximation where the particles' motion is non--relativistic and shown that this is still causal as it includes the corrections for retardation. As expected, retardation has an observable effect in the production of mediated entanglement. 

The  physical picture arising from our analysis is that the mechanism giving rise to entanglement is a quantum superposition of macroscopically distinct dynamical field configurations. Per equation \eqref{eq:phases}, it is this superposition that gives rise to different phases for each quantum branch.

Our analysis gives a complementary point of view to the work in \cite{marletto2018when,marshman2020locality,*bose2022mechanism,carney2021using,carney2021newton}, where it is concluded that the mediation of quantum information takes place through the exchange of virtual gravitons (or virtual photons). Indeed, at the level of perturbation theory, scattering potentials can be understood as the result of exchanging virtual particles. We have seen here that setting the field on shell (that is, neglecting quantum fluctuations) on each quantum branch is sufficient to recover the causal propagation of signals. A physical interpretation of our analysis is that quantum information propagates casually due to the field wavefronts being in a quantum superposition.

As an application of our results, we considered an experiment to detect retardedly induced entanglement. This is for the moment a gedanken experiment for gravity. Because of the theoretical and physical analogies, it is interesting to consider performing the analogous experiment in electromagnetism. We estimate this task to be challenging but plausible.

\subsubsection*{Acknowledgments}
    We acknowledge support of the ID\# 61466 and ID\# 62312 grants from the John Templeton Foundation, as part of the ``Quantum Information Structure of Spacetime (QISS)'' project (\hyperlink{http://www.qiss.fr}{qiss.fr}). CB acknowledges  support by the Austrian Science Fund (FWF) through BeyondC (F7103-N38),
    the European Commission via Testing the Large-Scale Limit of Quantum Mechanics (TEQ) (No. 766900) project, 
    and the Foundational Questions Institute (FQXi).
    M.A. and C.B. acknowledge support by the Austrian Academy of Sciences (OEAW) through the project ``Quantum Reference Frames for Quantum Fields'' (IF 2019 59 QRFQF). We thank Philipp Haslinger for helpful discussions, and Domenico Giulini for insightful and  helpful comments.
    
\appendix
\section{Appendix: Derivation of \eqref{eq:SosGravity}} \label{app:onshellaction}

Below we summarise the derivation of the on--shell action \eqref{eq:SosGravity} and explain the notation. A more pedagogical derivation is provided in the Supplementary Material, also for the electromagnetic case. 

The gauge--invariance of $S_{\mathcal{F}}$ can be used to simplify computations by writing the Lagrangian in the Lorenz gauge $
\partial^{\nu} \bar{h}_{\mu \nu} = 0$. 
The action for linearised gravity coupled to matter then simplifies to \cite{maggiore2008gravitational,carroll2004spacetime,poisson2014gravity,flanagan2005basics} 
\begin{align} \label{LorentzSG} 
	S_\mathcal{F} = &\frac{c^4}{64 \pi G} \int \dd^4 x \;  \left(-\partial_{\rho}  h_{\mu \nu} \partial^{\rho} h^{\mu \nu}  + \frac{1}{2} \partial^{\mu} h \partial_{\mu} h \right) \nonumber \\ &+ \frac{1}{2}\int d^4 x  h_{\mu \nu} T^{\mu \nu}.
\end{align} 
where $\dd^4x = \dd t \dd^3x$ and $T^\munu$ is the energy--momentum tensor. Greek indices denote 4--vectors and bold latin letters denote 3--vectors. The metric perturbation satisfies $|h_{\mu \nu}| \ll 1$ and $\eta_{\mu \nu}$ is the Minkowski metric. We  use the notation $X=\eta_\munu X^\munu$ for the trace and $\bar X^\munu = X^\munu - \frac12 \eta_\munu X$ the trace--reversed of a  2-tensor.

The Euler-Lagrange equations for the field are
\begin{align} \label{BoxhmunuG-supp}
	\Box h_{\mu \nu} = -\frac{16 \pi G}{c^4} \bar{T}_{\mu \nu}.
\end{align}
 When the field is taken on--shell, we can integrate by parts the terms with two derivatives of $h_\munu$ in \eqref{LorentzSG} to obtain two terms of the form $h_{\cdot\cdot}\square h^{\cdot\cdot}$ and use \eqref{BoxhmunuG-supp} to get \eqref{SLorentzG}.

Next, we consider the gravitational interaction of $N$ point particles of masses $m_a$. The use of point particles is an approximation that allows to use an explicit solution of the field equations. So long as the size of the two matter distributions is much smaller than their separation, so that finite size effects can be neglected, the use of point charges will be a good approximation.

The solution obtained here is the gravitational analogue of the well--known Li\'{e}nard–Wiechert potential of electromagnetism \cite{jackson1999classical, kopeikin1999lorentz}.

The stress--energy tensor for $N$ point masses is
 \begin{align} \label{TmunuG-supp} 
 	T^{\mu \nu} (t,\bm{x}) = \sum_{a=1}^N m_a  \delta^{(3)} (\bm{x} - \bm{x}_a(t)) V_a^{\mu \nu} (t)
 \end{align}
where
$
  V^{\mu \nu}_a(t) = \gamma_a (t) v_{a}^{\mu}(t) v_{a}^{ \nu}(t)$
with $v^{\mu}_a(t) = (c, \bm{v}_a )$, where $\bm{v}_a= d\bm x_a / dt$ is the velocity of particle $a$ and $\gamma_a(t) =  (1 - \vert\bm{v}_a(t)\vert^2/c^2)^{-1/2}$ the corresponding Lorentz factor.
The retarded solution of the wave equation \eqref{BoxhmunuG-supp} for all times is 
\begin{align} 
h_{\mu \nu}(t,\bm x) = \frac{4 G}{c^4} \int \dd x'^3\; \frac{\bar{T}_{\mu \nu} ( \bm{x}',t_r)}{\vert \bm{x}- \bm{x}'\vert},
\end{align}
with the retarded time $t_r=t_r(t,\bm x, \bm x')$ defined by $ct_r  = ct - \vert \bm{x}' - \bm{x}\vert $. 
Plugging in the expression for the energy--momentum tensor, we obtain
\begin{equation}
    h_\munu(t,\bm x) = \frac{4G}{c^4}\sum_a m_a \int d^3x'\frac{\delta^{(3)}\big(\bm x' - \bm x_a(t_r)\big)\bar V^\munu_a(t_r)}{|\bm x - \bm x'|}.
\end{equation}
To deal with the awkward dependence of the retarded time $t_r$ on $\bm x'$ we introduce an integration in a dummy time variable $t'$ over a delta function $\delta(t'- \tilde{t})$. We can then do the $\bm x'$ integration to get
\begin{align} 
h_{\mu \nu}(t, \bm x) = \frac{4 G}{c^4} \sum_a m_a \int \dd t' \frac{ \bar{V}_{\mu \nu} (t') \delta\big(t'-\tilde{t}(\bm x, t,t')\big)}{\vert \bm{x}- \bm{x}_a(t')\vert}.
\end{align}
For the remaining integration in $t'$, we make use of the identity
$\delta(f(y)) = \sum_i \frac{\delta(y-y_i)}{\vert \partial_yf(y_i) \vert}$ where $y_i$ are zeros of $f(y)$.
It follows that
\begin{equation}
\delta\big(t'-\tilde{t}(\bm x, t,t')\big) = \frac{\delta(t'-t_a)}{1 - \bm{d}_a \cdot \bm{v}_a(t_a)/(d_a c)},
\end{equation}
where the retarded time $t_a$ is implicitly defined as a function of $t$ and $\bm x$ as satisfying $c(t-t_a) = |\bm x - \bm x_a(t_a)|$. Here,
$t_a$ is the time at which the past lightcone of the event $(t,\bm x)$ intersects the worldline of particle $a$. We also defined the retarded displacement $\bm d_a = \bm{d}_a(t,\bm x) = \bm{x} - \bm{x}_a(t_a)$
and its magnitude $d_a = |\bm{d}_a|$.  One then obtains the following field
\begin{align}\label{LW-gravity-supp}
	h^{\mu \nu}(t,\bm{x}) =  \frac{4 G }{c^4} \sum_a   \frac{m_a \bar{V}^{\mu \nu}_a(t_a)}{d_a - \bm{d}_a \cdot \bm{v}_a(t_a)/c},
\end{align}
where the values of the field at any given spacetime point $(t,\bm x)$ depend exclusively on the behaviour of the particles on the past lightcone of $(t,\bm x)$.

Next we calculate the on--shell action of $N$ interacting point masses. We plug in the energy--momentum tensor \eqref{TmunuG-supp} for $N$ point particles into \eqref{SLorentzG} and perform the space integration
\begin{equation}
    S_{\mathcal{F}} = \frac14 \sum_{b=1}^N  \int \dd t\;  m_b V_b^\munu(t) h_\munu\big(t,x_b(t)\big).  
\end{equation}
Next, we use \eqref{LW-gravity-supp} to obtain \eqref{eq:SosGravity}
\begin{equation}\label{action-particles}
     S_{\mathcal{F}} = \frac{G}{c^4}\sum_{a,b}\int \dd t \frac{m_a m_b  \bar V_a^\munu(t_{ab}) V_{b\munu}(t)}{d_{ab}-\bm d_{ab}\cdot\bm v_{a}(t_{ab})/c}.
\end{equation}
We denote as $t_{ab}=t_{ab}(t)$ the retarded time, at which the past lightcone of the event $\big(t,\bm x_b(t)\big)$ intersects the timelike wordline of particle $a$. This is defined implicitly by
\begin{equation}
    c(t-t_{ab}) = |\bm x_b(t) - \bm x_a(t_{ab})|.
\end{equation}
We also defined the retarded displacement
\begin{equation}
\bm d_{ab} = \bm d_{ab}(t) = \bm x_b(t) - \bm x_a(t_{ab}), 
\end{equation}
and its magnitude $d_{ab} = |\bm d_{ab}|$.

\bibliography{references}

\onecolumngrid

\clearpage

\section{Supplementary Material}

\section{Detailed derivation of equation (8)}

Here, we give a pedagogical derivation of the on--shell action $S^\sigma_{\mathcal{F}}$ when the field $\mathcal{F}$ is the metric perturbation of linearised gravity sourced by point particles. The electromagnetic case proceeds similarly, see next section.

We  use the notation: given a two--indexed tensor $X^\munu$, $X=\eta_\munu X^\munu$ is the trace and $\bar X^\munu = X^\munu - \frac12 \eta_\munu X$ the trace--reversed tensor.

\subsubsection{Action for the on--shell field and arbitrary sources}

The action for linearised gravity coupled to matter is
\begin{equation} \label{supp:eq:actionGravity-supp}
\begin{aligned}
	S_{h} = \frac{c^4}{64 \pi G} \int \dd^4 x \Big( &-\partial_{\rho} h_{\mu \nu} \partial^{\rho} h^{\mu \nu} + 2 \partial_{\rho} h_{\mu \nu} \partial^{\nu} h^{\mu \rho} 
	- 2 \partial_{\nu} h^{\mu \nu} \partial_{\mu} h + \partial^{\mu} h \partial_{\mu} h \Big)
	+ \frac{1}{2} \int \dd^4 x \,h_{\mu \nu} T^{\mu \nu},
\end{aligned}
\end{equation}
where $\dd^4x = \dd t \dd^3x$ and $T^\munu$ is the energy--momentum tensor for matter. Boundary terms at infinity are taken to vanish. The coordinates $\bm x,t$ are standard Minkowski coordinates. Greek indices denote 4--vectors and bold latin letters denote 3--vectors.  The full spacetime metric is given by $g_{\mu\nu} = \eta_{\mu \nu} + h_{\mu \nu}$ with the metric perturbation satisfying $|h_{\mu \nu}| \ll 1$ and $\eta_{\mu \nu}$ the Minkowski metric. The  metric signature is $(-,+,+,+)$. The action $S_h$ is invariant under an infinitesimal change of coordinates $x^{\mu} \longrightarrow x^{\prime \mu} = x^{\mu} + \xi^{\mu} (x),$ under which the metric perturbation transforms as  $h^\prime_{\mu \nu} (x^\prime) = h_{\mu \nu}(x) - \partial_{\mu} \xi_{\nu}(x') - \partial_{\nu} \xi_{\mu}(x')$. 

We use the gauge--invariance of $S_h$ to simplify computations by writing the Lagrangian in Lorenz gauge. In this gauge, the field $h_\munu$ satisfies 
\begin{equation}\label{supp:eq:lorenz_gauge}
\partial^{\nu} \bar{h}_{\mu \nu} = 0
\end{equation}
and the action \eqref{supp:eq:actionGravity-supp} simplifies  to
\begin{equation} \label{supp:LorentzSG-supp} 
\begin{aligned} 
	S_h = &\frac{c^4}{64 \pi G} \int \dd^4 x \;  \big[-\partial_{\rho}  h_{\mu \nu} \partial^{\rho} h^{\mu \nu}  + \frac{1}{2} \partial^{\mu} h \partial_{\mu} h \big]
	+ \frac{1}{2}\int d^4 x  \; h_{\mu \nu} T^{\mu \nu}.
\end{aligned}
\end{equation}
The Euler-Lagrange equations for the field are
\begin{align} \label{supp:BoxhmunuG-supp}
	\Box h_{\mu \nu} = -\frac{16 \pi G}{c^4} \bar{T}_{\mu \nu}.
\end{align}
When the field is taken on--shell, we can integrate by parts the terms with two derivatives of $h_\munu$ in \eqref{supp:LorentzSG-supp} to obtain two terms of the form $h_{\cdot\cdot}\square h^{\cdot\cdot}$ and use \eqref{supp:BoxhmunuG-supp} to get
\begin{align} \label{supp:on-shell-lorenz-action-supp}
	S_{h} = \frac{1}{4} \int \dd^4 x \; h_{\mu \nu} T^{\mu \nu}.
\end{align}




\subsubsection{The field of $N$ point masses}

Let us now consider the gravitational interaction of $N$ point particles of masses $m_a$. The use of point particles is an approximation that allows to use an explicit solution of the field equations. So long as the size of the two matter distributions is much smaller than their separation, so that finite size effects can be neglected, the use of point charges will be a good approximation.

The solution obtained here is the gravitational analogue of the well--known Li\'{e}nard–Wiechert potential of electromagnetism \cite{jackson1999classical}. These solutions are already known, see for example \cite{kopeikin1999lorentz}. We include a full derivation here for completeness, as we are not aware of a published solution of the gravitational case.


The stress--energy tensor for $N$ point masses is
 \begin{align} \label{supp:TmunuG-supp} 
 	T^{\mu \nu} (t,\bm{x}) = \sum_{a=1}^N m_a  \delta^{(3)} (\bm{x} - \bm{x}_a(t)) V_a^{\mu \nu} (t)
 \end{align}
where
\begin{equation}
  V^{\mu \nu}_a(t) = \gamma_a (t) v_{a}^{\mu}(t) v_{a}^{ \nu}(t)
\end{equation}
with $v^{\mu}_a(t) = (c, \bm{v}_a )$, where $\bm{v}_a= d\bm x_a / dt$ is the velocity of particle $a$ and $\gamma_a(t) =  (1 - \vert\bm{v}_a(t)\vert^2/c^2)^{-1/2}$ the corresponding Lorentz factor.
The retarded solution of the wave equation \eqref{supp:BoxhmunuG-supp} for all times is 
\begin{align} 
h_{\mu \nu}(t,\bm x) = \frac{4 G}{c^4} \int \dd x'^3\; \frac{\bar{T}_{\mu \nu} ( \bm{x}',t_r)}{\vert \bm{x}- \bm{x}'\vert},
\end{align}
with the retarded time $t_r=t_r(t,\bm x, \bm x')$ defined by $ct_r  = ct - \vert \bm{x}' - \bm{x}\vert $. 
Plugging in the expression for the energy--momentum tensor, we obtain
\begin{equation}
    h_\munu(t,\bm x) = \frac{4G}{c^4}\sum_a m_a \int d^3x'\frac{\delta^{(3)}\big(\bm x' - \bm x_a(t_r)\big)\bar V^\munu_a(t_r)}{|\bm x - \bm x'|}.
\end{equation}
Before being able to perform the space integration, one needs to first eliminate the awkward dependence of the retarded time $t_r$ on $\bm x'$ by help of an integration in a dummy time variable $t'$ and a delta function:
\begin{align} 
h_{\mu \nu}(t, \bm x) = \frac{4 G}{c^4} \sum_a m_a \int \dd^3 x'  \int\dd t' \frac{ \bar{V}_{\mu \nu} ( t') \delta\big( \bm{x}' - \bm{x}_a(t')\big)  \delta(t'- \tilde{t})}{\vert \bm{x}- \bm{x}'\vert},
\end{align}
where $\tilde t = \tilde{t}(\bm x, t,t') = t - \vert \bm x - \bm x_a(t') \vert/c $. We can now do the $\bm x'$ integration to get
\begin{align} 
h_{\mu \nu}(t, \bm x) = \frac{4 G}{c^4} \sum_a m_a \int \dd t' \frac{ \bar{V}_{\mu \nu} (t') \delta\big(t'-\tilde{t}(\bm x, t,t')\big)}{\vert \bm{x}- \bm{x}_a(t')\vert}.
\end{align}
For the remaining integration in $t'$, we make use of the identity
\begin{equation}
\delta(f(y)) = \sum_i \frac{\delta(y-y_i)}{\vert \partial_yf(y_i) \vert},
\end{equation}
where $y_i$ are zeros of $f(y)$.
It follows that
\begin{equation}
\delta\big(t'-\tilde{t}(\bm x, t,t')\big) = \frac{\delta(t'-t_a)}{1 - \bm{d}_a \cdot \bm{v}_a(t_a)/(d_a c)},
\end{equation}
where the retarded time $t_a$ is implicitly defined as a function of $t$ and $\bm x$ as satisfying 
\begin{equation}\label{supp:eq:ta-supp}
c(t-t_a) = |\bm x - \bm x_a(t_a)|;
\end{equation}
$t_a$ is the time at which the past lightcone of the event $(t,\bm x)$ intersects the worldline of particle $a$. We also defined the retarded displacement
\begin{equation}\label{supp:eq:da-supp}
\bm d_a = \bm{d}_a(t,\bm x) = \bm{x} - \bm{x}_a(t_a),
\end{equation}
and its magnitude $d_a = |\bm{d}_a|$.  One then obtains the following field
\begin{align}\label{supp:LW-gravity-supp}
	h^{\mu \nu}(t,\bm{x}) =  \frac{4 G }{c^4} \sum_a   \frac{m_a \bar{V}^{\mu \nu}_a(t_a)}{d_a - \bm{d}_a \cdot \bm{v}_a(t_a)/c},
\end{align}
where the values of the field at any given spacetime point $(t,\bm x)$ depend exclusively on the behaviour of the particles on the past lightcone of $(t,\bm x)$.


\subsubsection{The action of $N$ interacting point masses}

Let us start by plugging in the energy--momentum tensor \eqref{supp:TmunuG-supp} for $N$ point particles  into \eqref{supp:on-shell-lorenz-action-supp} and performing the space integration:
\begin{equation}
    S_h = \frac14 \sum_{b=1}^N  \int \dd t\;  m_b V_b^\munu(t) h_\munu\big(t,x_b(t)\big).  
\end{equation}
Next, make use of the solution \eqref{supp:LW-gravity-supp} to obtain
\begin{equation}\label{supp:action-particles}
    S_h = \frac{G}{c^4}\sum_{a,b}\int \dd t \frac{m_a m_b  \bar V_a^\munu(t_{ab}) V_{b\munu}(t)}{d_{ab}-\bm d_{ab}\cdot\bm v_{a}(t_{ab})/c}.
\end{equation}
We denoted by $t_{ab}=t_{ab}(t)$ the time at which the past lightcone of the event $\big(t,\bm x_b(t)\big)$ intersects the timelike wordline of particle $a$, defined implicitly by
\begin{equation}
    c(t-t_{ab}) = |\bm x_b(t) - \bm x_a(t_{ab})|.
\end{equation}
We also defined the retarded displacement
\begin{equation}
\bm d_{ab} = \bm d_{ab}(t) = \bm x_b(t) - \bm x_a(t_{ab}), 
\end{equation}
and its magnitude $d_{ab} = |\bm d_{ab}|$. (Compare with equations \eqref{supp:eq:ta-supp} and \eqref{supp:eq:da-supp}).

When plugged in the formula (5) in the main text for the phases $\phi^\sigma=  S^\sigma_\mathcal{F} /\hbar$, one obtains (9). Note that the terms of the sum where $a=b$ are dropped, as they just contribute an overall (infinite) phase to the state.

\subsubsection{Energy-momentum conservation}

Note that the field \eqref{supp:LW-gravity-supp}, while certainly a solution of \eqref{supp:BoxhmunuG-supp}, is not necessarily the solution to the linearised Einstein field equations. That is because a solution of \eqref{supp:BoxhmunuG-supp} is also a solution to the Einstein field equations only if it also satisfies the gauge condition $\partial_\mu\bar h_\munu=0$. This in turn happens only if the energy momentum tensor is conserved, \textit{i.e.}, $\partial_\mu T^\munu$, which is not the case for arbitrary particle trajectories. 
To fix this, one should model the motion and energy momentum $\tilde T^\munu$ of the apparatus, so that the overall $T^\munu+\tilde T^\munu$ is conserved. This would contribute an additional term $\tilde h_\munu$ to the field, so that the total field $h_\munu+\tilde h_\munu$ satisfies the gauge constraint and thus is a solution to the linearised Einstein field equations.
One might then worry that computing the phases using only the stress energy tensor of the particles would lead to incorrect results.
However, the phases computed using the new overall energy-momentum tensor and field are the same as those computed using \eqref{supp:action-particles}, as long as the apparatus is heavy enough and that its field is homogeneous in the region traversed by the particles, which are reasonable assumptions. Indeed, the heaviness of the apparatus is what prevents interferometers and beamsplitters from becoming entangled with the particles traversing them, while the homogeneity of the field will be a function of the arrangement of the apparatus and can be ensured by proper design.

Let us see this in a little more detail. When considering the effect of the apparatus, two\footnote{Terms of the form $h_\munu \tilde T^\munu$ can be turned into terms of the form $\tilde h_\munu T^\munu$ by integrating by parts} more terms appear in \eqref{supp:on-shell-lorenz-action-supp}, one involving $\tilde h_\munu T^\munu$ (interaction between particles and apparatus) and one involving $\tilde h_\munu \tilde T^\munu$ (apparatus self-interaction). Now, the relevant quantities are the relative phases between the branches, and thus we are interested in \textit{differences} in the actions computed via \eqref{supp:on-shell-lorenz-action-supp}. In each branch, the apparatus has a different motion to account for the change in momentum of the particles. But if the apparatus is heavy enough, it will not move significantly, and thus the term $\tilde h_\munu \tilde T^\munu$ will be pretty much the same in each branch, and thus not contribute to the relative phases.  The term $\tilde h_\munu T^\munu$ will similarly drop out if the field is homogeneous enough in the volume traversed by the various particle trajectories. This can be verified by doing a computation in the Newtonian regime.

\section{Equations (8) and (9) for Electromagnetism}

The action for electromagnetism coupled to a four current $j^\mu$ is of the form $S=S_M+S_A$ with
\begin{align}
	S_A =    \int \dd^4 x \left(-\frac{c^2}{16 \pi k_e} F_{\mu \nu} F^{\mu \nu} +  A_{\mu}j^\mu\right).
\end{align}
 Here, $d^4 x = dt d^3x$, $S_M$ is the free matter action that also includes the $B_0$ coupling to the spins,  $A_{\mu} $ is the four-potential,  $F_{\mu \nu} = \partial_{\mu} A_{\nu} - \partial_{\nu} A_{\mu}$ is the field strength and $k_e$ Coulomb's constant. We use greek indexed latin letters for 4--vectors and bold latin letters for 3--vectors. The  metric signature is $(-,+,+,+)$ and $\eta_{\mu \nu}$ is the Minkowski metric.  

The action $S_A$ is gauge invariant. We will express the Lagrangian in the Lorentz gauge $\partial_{\mu} A^{\mu} = 0$ to simplify calculations. 
Boundary terms at infinity are taken to vanish.  Integrating by parts, the action then reduces to 
\begin{align} \label{supp:SLorentzEM-supp}
	S_A = \int \dd^4 x \left(- \frac{c^2}{8 \pi k_e} \partial_{\mu} A_{\nu} \partial^{\mu} A^{\nu} + j_{\mu} A^{\mu} \right),
\end{align}
and the equations of motion are
\begin{align} \label{supp:EqnsMotionLorentz-supp}
	\Box A^{\mu} = -\frac{4 \pi k_e}{c^2} j^{\mu},
\end{align}
where $\Box =\partial^{\mu} \partial_{\mu}$.
Now, we obtain the on--shell action by integrating again  \eqref{supp:SLorentzEM-supp} by parts and using \eqref{supp:EqnsMotionLorentz-supp} to get: 
\begin{equation}
\label{supp:eq:osintAction-supp}
S^\sigma_{A} = \frac{1}{2} \int \dd^4x\;  j_{\mu} A^{\mu}.
\end{equation}
The entire contribution of the electromagnetic field to the on--shell action is encoded in this expression. As we saw in the main text, $S^\sigma_{A}$ is the central object of interest for mediated entanglement: this Lorentz covariant and gauge invariant quantity is the observable that would be measured in an experiment aiming to observe mediated entanglement.

We now consider the electromagnetic interaction of $N$ point charges. The four-current is given by
\begin{align}\label{supp:point_current-supp}
	j^{\mu}(x) = \sum_a q_a v_a^{\mu}(t) \delta^{(3)} (\bm{x} - \bm{x}_a (t) ),
	\end{align}	
where $v^{\mu}_a(t) = (c, d \bm{x}_{a} / dt)= (c, \bm{v}_{a})$.
The potential of this charge configuration in the Lorenz gauge is the well--known Li\'{e}nard–Wiechert potential \cite{jackson1999classical}
\begin{align}
\label{supp:eq:LWpotential-supp}
	A^{\mu}(t,\bm{x}) = \frac{k_e }{c^2 } \sum_a  \frac{q_a v^{\mu}_a(t_a) }{d_a - \bm{d}_a \cdot \bm{v}_a(t_a)/c},
\end{align}	
where $t_a$ is the \emph{retarded} time, defined implicitly as the solution of
\begin{align}
  c(t-t_a) = |\bm x - \bm x_a(t_a)|.
\end{align}
The retarded time $t_a$ is the time at which the worldine of particle $a$ intersects the past light-cone of $(t,\bm x)$. We also defined, for convenience, the retarded displacement $ \bm d_a = \bm d_a(t,\bm x) =\bm{x}-\bm{x}_a(t_a)$, and its magnitude $d_a=\vert \bm{d_a} \vert$. 

By placing \eqref{supp:point_current-supp} and \eqref{supp:eq:LWpotential-supp} in \eqref{supp:eq:osintAction-supp} and performing the space integration, we get an explicit expression for the on--shell action giving the interaction between the two charges
\begin{align}\label{supp:Sint00exactEM-supp}
	S^\sigma_{A} & = \frac{k_e }{2 c^2} \sum^{a \neq b}_{a,b} \int \dd t  \;    \frac{q_a\, q_b\, v^{\mu}_a(t_{ab}) \, v_{b \mu} (t)}{d_{ab} - \bm{d}_{ab} \cdot \bm{v}_a(t_{ab})/c} .
\end{align}	
Here, the retarded time $t_{ab}$ is defined as the implicit solution of
\begin{equation}
c(t-t_{ab}) = |\bm x_b(t) - \bm x_a(t_{ab})|,
\end{equation}
and we also defined $\bm{d}_{ab} = \bm d_{ab}(t) = \bm x_{b}(t) - \bm x_a(t_{ab})$, and $d_{ab} = |\bm d_{ab}|$.

In the slow moving approximation $|\bm{v}_a| \ll c $ the exact expression \eqref{supp:Sint00exactEM-supp} approximates to the retarded Coulomb interaction
\begin{align} \label{supp:Sint00EM-supp}
	S^\sigma_{\mathcal{F}} \approx -\frac{1}{2} k_e \sum_{a,b}^{a \neq b}  \int \dd t\;  \frac{q_a q_b}{d_{ab}}.
\end{align}

\end{document}